\documentclass[a4paper,11pt]{article}
\pdfoutput=1 % if your are submitting a pdflatex (i.e. if you have
\usepackage{jcappub} % for details on the use of the package, please
\usepackage[T1]{fontenc} % if needed

\title{\boldmath Reviving Quadratic Inflation: Minimal Deformation for CMB Compatibility and Reheating Consistency}

\author[a,1]{K. Djeha,\note{Corresponding author.}}
\author[b]{Y. Derouiche}
\author[c]{ M. Hadj Moussa }

\affiliation[a]{Physique Théorique et Interactions Rayonnement-Matière, Blida University, P.O.Box 270, Route de Soumaa, Blida 0900, Algeria}
\affiliation[b]{Laboratoire Physico-Chimique des Matériaux et Environnement (LPCME), Faculty of Exact Sciences and Computer Science, University of Djelfa, P.O. Box 3117, Djelfa 17000, Algeria}
\affiliation[c]{Physique Théorique et Interactions Rayonnement-Matière, Blida University, P.O.Box 270, Route de Soumaa, Blida 0900, Algeria}
\emailAdd{djeha\_khadidja@univ-blida.dz}
  
\abstract{
We revisit the quadratic inflationary potential by introducing a minimal 
higher-order correction obtained through a simple field redefinition, 
leading to 
\[
V(\chi) = \tfrac{1}{2} m^2 \left( \chi - \tfrac{\gamma}{14}\chi^7 \right)^2.
\] 
While the uncorrected quadratic model predicts 
$(n_s \simeq 0.967,\, r \simeq 0.13)$, in strong tension with CMB data, 
the corrected potential yields 
$(n_s \simeq 0.965,\, r \simeq 0.036)$, 
fully consistent with Planck 2018 constraints. 

Beyond inflationary observables, the deformation also impacts the reheating phase. 
In the quadratic case reheating corresponds to a matter-like regime with 
$w_{\rm reh}=0$, whereas the corrected potential gives 
$w_{\rm reh}\simeq -1.1\times10^{-2}$, a slightly softer equation of state. 
This modification raises the reheating temperature by a factor of $\sim 3.4$ 
(for $N_{\rm reh}=10$), or equivalently extends the reheating duration at fixed 
temperature. 

Our results demonstrate that even a minimal higher-order correction is sufficient 
to reconcile the quadratic model with observations while simultaneously providing 
a more consistent post-inflationary history, highlighting the relevance of 
controlled deformations of simple inflationary scenarios.}

\begin{document}
\maketitle
\flushbottom
\section{Introduction} 
\label{sec:intro}
Inflationary cosmology has become an essential paradigm in modern theoretical physics, offering a compelling resolution to the classical shortcomings of the standard Big Bang scenario, including the horizon, flatness, and monopole problems. In particular, the horizon problem—why causally disconnected regions of the Universe exhibit nearly identical temperatures—finds a natural solution via a brief phase of accelerated expansion in the early Universe, during which causal contact is established across large comoving scales.

In addition to addressing these theoretical puzzles, inflation provides a robust mechanism for generating primordial curvature perturbations, whose nearly scale-invariant spectrum is imprinted on the cosmic microwave background (CMB). These predictions have been tested with high precision by recent observations, especially those of the \textit{Planck} satellite~\cite{planck2020}, which constrain the scalar spectral index to \( n_s = 0.9649 \pm 0.0042 \) and place a stringent upper limit on the tensor-to-scalar ratio, \( r < 0.056 \) (95\% CL). These data strongly disfavor simple monomial potentials such as the quadratic model \( V(\phi) \propto \phi^2 \), and motivate the search for theoretically consistent models that predict lower values of \( r \) while remaining consistent with the observed \( n_s \).

One well-explored route is to invoke modifications of gravity or to employ non-minimal couplings, as in the Starobinsky model~\cite{starobinsky1980}, Higgs inflation~\cite{bezrukov2008}, or supergravity-based $\alpha$-attractors~\cite{kallosh2013,roest2014}. Alternatively, inflationary models with non-canonical kinetic terms have received increasing attention, as they can effectively flatten the potential in the canonical frame without modifying the underlying gravitational theory. Notable examples include K-inflation~\cite{armendarizpicon2001}, DBI inflation~\cite{alishahiha2004}, and models involving non-trivial field-space geometries~\cite{garriga1999}.

In this work, we investigate a minimal single-field inflationary scenario governed by a non-canonical kinetic structure of the form
\begin{equation}
K(\phi) = 1 + \gamma \phi^6,
\end{equation}
where \( \gamma \) is a small positive coupling parameter. The scalar potential is taken to be purely quadratic:
\begin{equation}
V(\phi) = \frac{1}{2} m^2 \phi^2,
\end{equation}
but due to the field redefinition required to move to the canonical basis, the effective potential is dynamically flattened. At leading order, the canonically normalized field \( \chi \) yields an effective potential of the form
\begin{equation}
\tilde{V}(\chi) = \frac{1}{2} m^2 \chi^2 - \frac{1}{14} m^2 \gamma \chi^8 + \mathcal{O}(\chi^{14}),
\end{equation}
which features suppressed tensor modes and slightly redder spectral tilt compared to the original quadratic model.

The novelty of this approach lies in the fact that the flattening is not imposed arbitrarily, but emerges naturally from the kinetic term structure of a well-defined Lagrangian. This enhances the theoretical consistency of the model and its connection to field-theoretic foundations.

We perform a detailed analysis of the inflationary dynamics, including the evolution of slow-roll parameters, the number of e-folds, and the scalar and tensor power spectra. We show that the model yields predictions for \( (n_s, r) \) consistent with the 68\% confidence region of \textit{Planck} and BICEP/Keck data, with typical values such as \( n_s \simeq 0.964 \) and \( r \simeq 0.033 \) for \( N \simeq 59 \). We also comment on the role of the reheating phase and its impact on the inflationary predictions.

The paper is structured as follows: In Section~\ref{sec:theory}, we present the theoretical setup, including a non-canonical scalar field Lagrangian, and derive the canonically normalized potential using a perturbative field redefinition. We also discuss the physical implications of the resulting effective potential and its relevance to inflationary dynamics. In Section~\ref{sec:numerical}, we analyze the inflationary predictions of the model numerically and confront them with recent CMB observational data. Section~\ref{sec:reheating} briefly explores the implications for reheating in this framework. Finally, our conclusions and future directions are summarized in Section~\ref{sec:conclusion}.

\section{Theoretical Setup}\label{sec:theory}

We consider a scalar field theory with a non-canonical kinetic term, described by the Lagrangian:
\begin{equation}
\mathcal{L} = \frac{1}{2} K(\phi) \partial_\mu \phi \partial^\mu \phi - V(\phi),
\end{equation}
where the kinetic function and potential are given by:
\begin{equation}
K(\phi) = 1 + \gamma \phi^6, \quad V(\phi) = \frac{1}{2} m^2 \phi^2,
\end{equation}
with \(\gamma\) a small, positive coupling constant.

The choice \(K(\phi) = 1 + \gamma \phi^6\) represents the leading-order, symmetry-preserving deviation from a canonical kinetic term, consistent with the discrete \(\mathbb{Z}_2\) symmetry \(\phi \to -\phi\). Such higher-dimensional operators commonly arise in effective field theory (EFT) via radiative corrections or upon integrating out heavy fields~\cite{burgess2007, georgi1993}. The \(\phi^6\) contribution is the lowest-order non-trivial term consistent with analyticity and symmetry.

To transform the kinetic term into canonical form, we redefine the field via:
\begin{equation}
\frac{d\chi}{d\phi} = \sqrt{1 + \gamma \phi^6}.
\end{equation}
This approach is widely used in models involving non-canonical or k-inflationary frameworks~\cite{armendarizpicon2001, garriga1999}.

Expanding perturbatively for \(\gamma \ll 1\), integrating, and inverting the series yields:
\begin{equation}
\phi = \chi - \frac{1}{14} \gamma \chi^7 + \text{higher-order terms}.
\end{equation}

A detailed derivation of this result, including second-order terms in \(\gamma\), is provided in Appendix~\ref{app:canonical_potential}. There, we show that substituting the inverse relation into the original potential results in the canonical form:
\begin{equation}
\tilde{V}(\chi) = \frac{1}{2} m^2 \chi^2 - \frac{1}{14} m^2 \gamma \chi^8 + \text{higher-order corrections}.
\end{equation}

We verify numerically that truncating the expansion at leading order in \(\gamma\) provides excellent accuracy for observationally viable values of \(\gamma\). Subleading terms, such as the \(\chi^{14}\) correction, modify inflationary observables negligibly and are therefore omitted in the main analysis for clarity.

The canonical potential exhibits a flattening at large field values due to the \(-\chi^8\) term. This flattening dynamically reduces the inflaton’s speed during the slow-roll phase, leading to a suppressed tensor-to-scalar ratio \(r\), in better agreement with recent CMB observations~\cite{planck2020, bicep2021}, while preserving the simplicity of a quadratic potential in the original frame~\cite{linde1983}.

Although our focus here is on the background dynamics and inflationary observables, we briefly comment on the dynamical stability of the inflationary solution. A full stability analysis would involve perturbing the system and deriving the second-order action for fluctuations~\cite{garriga1999}. While this is beyond the scope of the present work, the monotonicity and smoothness of the canonical potential, together with the fulfillment of standard slow-roll conditions, suggest that the inflationary trajectory is stable within the field range of interest. A more rigorous stability treatment, including both scalar and tensor perturbations, is deferred to future work.
\section{Inflationary Dynamics and Numerical Methods}
\label{sec:dynamics}

\subsection{Background Equations of Motion}

We consider a spatially flat Friedmann–Lemaître–Robertson–Walker (FLRW) universe with the line element:
\begin{equation}
ds^2 = -dt^2 + a(t)^2 d\vec{x}^2,
\end{equation}
where \(a(t)\) is the scale factor. The dynamics of the homogeneous scalar field \(\chi(t)\) minimally coupled to gravity is governed by the following equations:
\begin{align}
H^2 &= \frac{1}{3 M_{\text{Pl}}^2} \left( \frac{1}{2} \dot{\chi}^2 + \tilde{V}(\chi) \right), \label{eq:Friedmann} \\
\ddot{\chi} + 3H\dot{\chi} + \frac{d\tilde{V}}{d\chi} &= 0, \label{eq:KleinGordon}
\end{align}
where \(H = \dot{a}/a\) is the Hubble parameter, and \(M_{\text{Pl}}\) is the reduced Planck mass. The potential \(\tilde{V}(\chi)\) is the canonical potential derived previously in Section~\ref{sec:theory}. These background equations are standard in inflationary cosmology~\cite{linde1983, mukhanov2005, baumann2009}.

\subsection{Numerical Methodology}

We solve the coupled system of equations~\eqref{eq:Friedmann} and~\eqref{eq:KleinGordon} numerically using standard finite-difference methods. The system is evolved in cosmic time \(t\), with initial conditions chosen in the slow-roll regime, where the kinetic energy is subdominant compared to the potential energy:
\begin{equation}
\dot{\chi}_\text{init} \ll \sqrt{2\tilde{V}(\chi_\text{init})}.
\end{equation}

We implement the numerical integration using a Runge-Kutta (RK4) algorithm, with adaptive step size control to ensure accuracy during the rapid field evolution stages.

The key inflationary observables—such as the number of e-folds \(N\), the slow-roll parameters, and the field excursion—are computed as functions of time throughout the inflationary phase. In particular, the number of e-folds is given by:
\begin{equation}
N = \int_{t_{\text{init}}}^{t_{\text{end}}} H(t) dt,
\end{equation}
where \(t_{\text{end}}\) is determined dynamically when the first slow-roll parameter satisfies \(\epsilon = -\dot{H}/H^2 = 1\)~\cite{lyth1999}.

\subsection{Initial Conditions and Assumptions}

To ensure a physically relevant inflationary trajectory, we choose initial field values \(\chi_\text{init}\) large enough such that slow-roll conditions are satisfied:
\begin{equation}
\epsilon \ll 1, \quad \eta = \frac{1}{M_{\text{Pl}}^2} \frac{V''}{V} \ll 1.
\end{equation}

We assume the universe starts in a classical homogeneous state, with negligible initial radiation or matter content. The value of the coupling parameter \(\gamma\) is chosen to be small (\(\gamma \ll 1\)) so that the perturbative expansion remains valid. All physical quantities are expressed in reduced Planck units (\(M_{\text{Pl}} = 1\)) unless otherwise stated.

\section{Numerical Analysis of Cosmological Parameters}\label{sec:numerical}

The numerical analysis of the inflationary parameters was carried out based on the following scalar potential model:
\[
V(\chi) = \frac{1}{2} m^2 \left( \chi - \frac{\gamma}{14} \chi^7 \right)^2,
\]
where \( \chi \) denotes the inflaton field, and \( m \), \( \gamma \) are model parameters chosen within small values consistent with observational constraints~\cite{planck2020}.

To derive analytical expressions for the slow-roll parameters, we define the function
\[
f(\chi) = \chi - \frac{\gamma}{14} \chi^7,
\]
so that the potential becomes \( V(\chi) = \frac{1}{2} m^2 f(\chi)^2 \). Differentiating, we find:
\begin{align*}
f'(\chi) &= 1 - \frac{\gamma}{2} \chi^6, \\
f''(\chi) &= -3\gamma \chi^5.
\end{align*}

This leads to the following expressions for the slow-roll parameters:
\begin{align}
\epsilon(\chi) &= \frac{2 \left( f'(\chi) \right)^2}{\left( f(\chi) \right)^2}
= \frac{2 \left( 1 - \frac{\gamma}{2} \chi^6 \right)^2}{\left( \chi - \frac{\gamma}{14} \chi^7 \right)^2}, \label{eq:epsilon_analytic} \\
\eta(\chi) &= \frac{2}{f(\chi)^2} \left[ f'(\chi)^2 + f(\chi) f''(\chi) \right]
= \frac{2 \left[ \left( 1 - \frac{\gamma}{2} \chi^6 \right)^2 - 3\gamma \chi^5 \left( \chi - \frac{\gamma}{14} \chi^7 \right) \right]}{\left( \chi - \frac{\gamma}{14} \chi^7 \right)^2}. \label{eq:eta_analytic}
\end{align}

The scalar spectral index \( n_s \) and tensor-to-scalar ratio \( r \) are then obtained using the standard slow-roll relations~\cite{lidsey1997, baumann2009}:
\[
n_s = 1 - 6\epsilon + 2\eta, \quad r = 16\epsilon.
\]

Additionally, the number of e-folds is evaluated as:
\begin{equation}
N(\chi) = \int_{\chi_\text{end}}^{\chi} \frac{V(\chi)}{V'(\chi)} \, d\chi = \int_{\chi_\text{end}}^{\chi} \frac{f(\chi)}{f'(\chi)} \, d\chi = \int_{\chi_\text{end}}^{\chi} \frac{\chi - \frac{\gamma}{14} \chi^7}{1 - \frac{\gamma}{2} \chi^6} \, d\chi. \label{eq:N_analytic}
\end{equation}

All numerical calculations were performed using \texttt{Mathematica}, based on symbolic differentiation and adaptive numerical integration. The analysis was carried out at \( \chi = 19.5 \), \( \chi_\text{end} = 1 \), with parameter choices \( m = 0.001 \), and \( \gamma = 1.5 \times 10^{-8} \).

Based on this setup, the resulting values of the cosmological parameters are summarized in Table~\ref{tab:results}.

\begin{table}[tbp]
\centering
\begin{tabular}{|lr|c|}
\hline
\textbf{Parameter} & & \textbf{Value} \\
\hline
\( \chi \) & & 19.5 \\
\( \chi_{\text{end}} \) & & 13.66 \\
\( m \) & & 0.001 \\
\( \gamma \) & & \( 1.5 \times 10^{-8} \) \\
\( V(\chi) \) & & 0.000168385 \\
\( \epsilon(\chi) \) & & 0.00205083 \\
\( \eta(\chi) \) & & -0.0117768 \\
\( n_s \) & & 0.964141 \\
\( r \) & & 0.0328133 \\
Inflation (\( \epsilon < 1 \))? & & Yes \\
Number of e-folds \( N \) & & 59.1377 \\
\hline
\end{tabular}
\caption{\label{tab:results} Numerical results of inflationary parameters based on the deformed quadratic potential.}
\end{table}

\section*{Observational Predictions}

Using the numerical results summarized in Table~\ref{tab:results}, 
we compute the observational parameters of interest, namely the 
spectral index \( n_s \) and the tensor-to-scalar ratio \( r \). 
These quantities are then compared with the latest CMB constraints 
from Planck~\cite{planck2020}. In particular, the trajectory in the 
\( n_s - r \) plane predicted by the deformed quadratic potential 
(\( \gamma \neq 0 \)) is contrasted with the standard quadratic model 
(\( \gamma = 0 \)). 

\begin{figure}[tbp]
\centering
\includegraphics[width=0.95\textwidth]{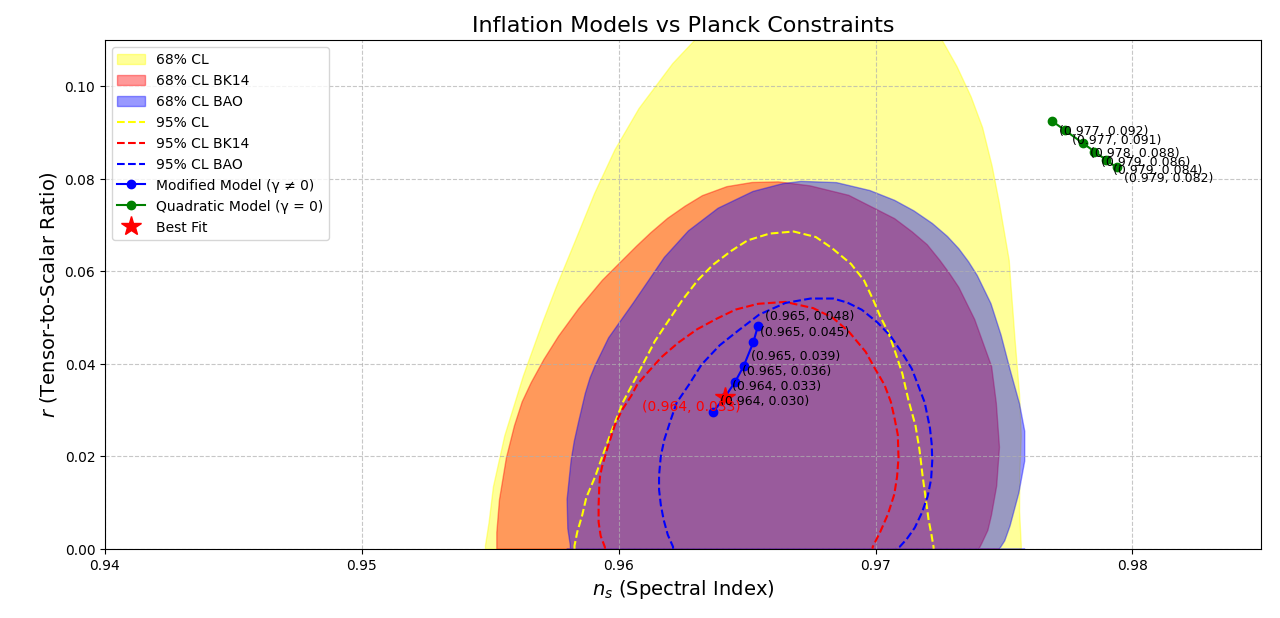}
\caption{\label{fig:ns_r_plot}
Comparison between the observational constraints from Planck, Planck+BK14, 
and Planck+BAO at 68\% and 95\% confidence levels (colored contours), 
and the theoretical predictions in the \( n_s - r \) plane. 
The green points correspond to the standard quadratic model (\( \gamma = 0 \)), 
which lies outside the observationally allowed region. 
In contrast, the blue trajectory corresponds to the modified model 
(\( \gamma \neq 0 \)), which falls inside the 68\% CL region and 
approaches the best-fit point (red star). 
This highlights the improved consistency of the deformed quadratic potential 
with current cosmological observations.
}
\end{figure}

The results are displayed in Fig.~\ref{fig:ns_r_plot}, where the 
confidence contours at 68\% and 95\% CL are also shown for Planck, 
Planck+BK14, and Planck+BAO datasets. 

From the figure, we observe that the standard quadratic model 
(\( \gamma = 0 \)) predicts values of \( n_s \) and \( r \) that 
lie outside the 95\% confidence region, hence being disfavored by 
observations. In contrast, the modified model with a small 
deformation parameter (\( \gamma \neq 0 \)) shifts the predictions 
significantly towards the allowed region. The resulting trajectory 
enters well inside the 68\% confidence level and approaches the 
best-fit point, highlighting the efficiency and relevance of the 
proposed correction to the potential. 

This demonstrates that the inclusion of the deformation parameter 
provides a better agreement with observational data and offers a 
viable alternative to the standard quadratic scenario.

\section{Reheating Dynamics}
\label{sec:reheating}
The reheating stage is sensitive to the inflaton potential near its minimum. 
For the quadratic model, $V(\chi)=\tfrac12 m^2\chi^2$, oscillations are harmonic and
\[
w_{\rm reh}^{\rm (quad)}\simeq 0,
\]
leading to the standard relations
\[
N_{\rm reh}^{\rm (quad)}=\frac{1}{3}\ln\!\left(\frac{V_{\rm end}}{\tfrac{\pi^2}{30}g_*\,T_{\rm reh}^4}\right),
\qquad
T_{\rm reh}^{\rm (quad)}=\left(\frac{30}{\pi^2 g_*}V_{\rm end}\right)^{\!1/4}
e^{-\tfrac{3}{4}N_{\rm reh}}.
\]

For the deformed potential studied here,
\[
V(\chi)=\tfrac12 m^2\!\left(\chi-\tfrac{\gamma}{14}\chi^7\right)^2,
\]
the higher-order term slightly modifies the oscillatory dynamics. Expanding near the minimum,
$V(\chi)\approx \tfrac12 m^2\chi^2+\alpha\,\chi^8$ with $\alpha=-m^2\gamma/14<0$, one finds a
small negative correction to the effective equation of state,
\[
w_{\rm reh}^{\rm (def)}\simeq -\frac{15}{128}\,\gamma\,A^6,
\]
where $A$ is the oscillation amplitude at the end of inflation (we take $A\simeq\chi_{\rm end}$).
Accordingly,
\[
N_{\rm reh}^{\rm (def)}=\frac{1}{3\!\left(1+w_{\rm reh}^{\rm (def)}\right)}
\ln\!\left(\frac{V_{\rm end}}{\tfrac{\pi^2}{30}g_*\,T_{\rm reh}^4}\right),
\qquad
T_{\rm reh}^{\rm (def)}=\left(\frac{30}{\pi^2 g_*}V_{\rm end}\right)^{\!1/4}\!
\exp\!\left[-\tfrac{3}{4}\!\left(1+w_{\rm reh}^{\rm (def)}\right)N_{\rm reh}\right].
\]

\paragraph{Numerical inputs and end-of-inflation energies.}
Using $m=10^{-3}$, $\gamma=1.5\times 10^{-8}$, $g_*=106.75$, and $M_{\rm Pl}=2.435\times10^{18}\,\mathrm{GeV}$:
\[
V_{\rm end}^{\rm (quad)}=\tfrac12 m^2 \chi_{\rm end}^2\Big|_{\epsilon=1}
=\tfrac12 m^2\,(2)=m^2=10^{-6},
\]
\[
\chi_{\rm end}^{\rm (def)}\approx 13.66\;\Rightarrow\;
f(\chi_{\rm end})=\chi_{\rm end}-\frac{\gamma}{14}\chi_{\rm end}^7\simeq 13.565,
\quad
V_{\rm end}^{\rm (def)}=\tfrac12 m^2 f(\chi_{\rm end})^2 \simeq 9.20\times 10^{-5}.
\]
With $A\simeq \chi_{\rm end}$, the deformed-model equation of state is
\[
w_{\rm reh}^{\rm (def)}\simeq -\frac{15}{128}\,\gamma\,A^6
\approx -1.14\times 10^{-2}.
\]

\paragraph{Scenario A (fixed $N_{\rm reh}=10$):}
The reheating temperatures are
\[
T_{\rm reh}^{\rm (quad)}=\left(\frac{30}{\pi^2 g_*}V_{\rm end}^{\rm (quad)}\right)^{\!1/4}
e^{-\tfrac{3}{4}\cdot 10}\;\simeq\;7.18\times 10^{-6}\;M_{\rm Pl}
\;\;\approx\;1.75\times 10^{13}\ \mathrm{GeV},
\]
\[
T_{\rm reh}^{\rm (def)}=\left(\frac{30}{\pi^2 g_*}V_{\rm end}^{\rm (def)}\right)^{\!1/4}
\exp\!\left[-\tfrac{3}{4}(1+w_{\rm reh}^{\rm (def)})\cdot 10\right]
\simeq 2.42\times 10^{-5}\;M_{\rm Pl}
\;\;\approx\;5.90\times 10^{13}\ \mathrm{GeV}.
\]
Thus, for the same reheating duration, the deformed model yields a higher $T_{\rm reh}$.

\paragraph{Scenario B (fixed $T_{\rm reh}=10^{9}\ \mathrm{GeV}$):}
Converting $T_{\rm reh}$ to Planck units via $T/M_{\rm Pl}$, one finds
\[
N_{\rm reh}^{\rm (quad)}=\frac{1}{3}\ln\!\left(\frac{V_{\rm end}^{\rm (quad)}}{\tfrac{\pi^2}{30}g_*\,T^4}\right)
\simeq 23.03,
\qquad
N_{\rm reh}^{\rm (def)}=\frac{1}{3(1+w_{\rm reh}^{\rm (def)})}
\ln\!\left(\frac{V_{\rm end}^{\rm (def)}}{\tfrac{\pi^2}{30}g_*\,T^4}\right)
\simeq 24.82.
\]
Hence, for the same target temperature, the deformed model requires a longer reheating phase.

\paragraph{Takeaway.}
Because $w_{\rm reh}^{\rm (def)}<0$ slightly and $V_{\rm end}^{\rm (def)}$ is larger for the parameters of interest,
the deformed potential either achieves a higher reheating temperature at fixed $N_{\rm reh}$ or, equivalently,
requires a larger $N_{\rm reh}$ at fixed $T_{\rm reh}$, compared to the quadratic case.
\begin{table}[h!]
\centering
\caption{Comparison of reheating parameters between the quadratic and deformed potentials. 
Units: $M_{\rm Pl}=2.435\times10^{18}\,\mathrm{GeV}$.}
\label{tab:reh}
\begin{tabular}{lccc}
\hline\hline
 & $w_{\rm reh}$ & $T_{\rm reh}$ [GeV] & $N_{\rm reh}$ \\
\hline
Quadratic ($V=\tfrac12 m^2\chi^2$) 
& $\;0$ 
& $1.75\times 10^{13}$ (for $N_{\rm reh}=10$) 
& $23.0$ (for $T_{\rm reh}=10^{9}$) \\
Deformed ($V=\tfrac12 m^2(\chi-\tfrac{\gamma}{14}\chi^7)^2$) 
& $-1.14\times 10^{-2}$ 
& $5.90\times 10^{13}$ (for $N_{\rm reh}=10$) 
& $24.8$ (for $T_{\rm reh}=10^{9}$) \\
\hline
\end{tabular}
\end{table}
\subsection{Comparison of Reheating Dynamics}

A quantitative comparison between the quadratic potential and the deformed model
highlights the impact of the $\gamma$-deformation on the post-inflationary phase:

\begin{itemize}
  \item \textbf{Equation-of-state parameter.}  
  For the quadratic case one finds $w_{\rm reh}=0$, corresponding to coherent
  harmonic oscillations of the inflaton around the minimum.  
  In contrast, the deformed model yields $w_{\rm reh}\simeq -1.1\times 10^{-2}$,
  indicating slightly softer dynamics with a small negative pressure on average.

  \item \textbf{Reheating temperature at fixed duration.}  
  For $N_{\rm reh}=10$, the quadratic model gives 
  $T_{\rm reh}\simeq 1.8\times 10^{13}\,\mathrm{GeV}$,
  while the deformed potential predicts a significantly higher value 
  $T_{\rm reh}\simeq 5.9\times 10^{13}\,\mathrm{GeV}$,
  i.e.\ an enhancement by a factor of about $3.4$.

  \item \textbf{Duration at fixed reheating temperature.}  
  For $T_{\rm reh}=10^{9}\,\mathrm{GeV}$, the quadratic case corresponds to
  $N_{\rm reh}\simeq 23.0$, whereas the deformed potential requires
  $N_{\rm reh}\simeq 24.8$, i.e.\ an extension of the reheating phase
  by almost $2$ e-folds.

  \item \textbf{Physical interpretation.}  
  The origin of these differences lies in the larger inflationary energy scale 
  at the end of inflation, $V_{\rm end}$, and in the negative correction to 
  $w_{\rm reh}$ introduced by the $\gamma$ term. The latter reduces the 
  effective dilution rate of the energy density during oscillations, leading to 
  either a higher reheating temperature for the same duration, or a longer 
  reheating stage for the same temperature.

  \item \textbf{Cosmological implications.}  
  Both models are easily consistent with BBN bounds since 
  $T_{\rm reh}\gg \mathcal{O}(\mathrm{MeV})$. 
  However, the deformed potential provides a more flexible and physically 
  appealing post-inflationary history: it not only improves the agreement of 
  inflationary observables $(n_s,r)$ with CMB constraints, but also shifts the 
  reheating regime towards values better aligned with cosmological bounds 
  from BBN and large-scale structure.
\end{itemize}
\begin{figure}[tbp]
\centering
\includegraphics[width=0.9\textwidth]{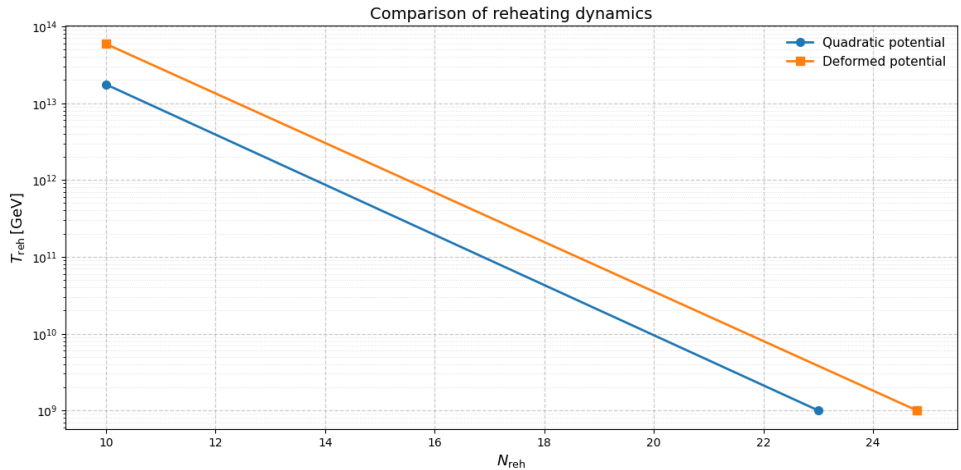}
\caption{\label{fig:Nreh_Treh}
Comparison of reheating dynamics in quadratic and deformed potentials, 
showing the relation between $N_{\rm reh}$ and $T_{\rm reh}$.
}
\end{figure}
As shown in Fig.~\ref{fig:Nreh_Treh}, the deformed potential leads to higher reheating temperature...
\section{Conclusion and Outlook}
\label{sec:conclusion}

We have proposed and analyzed a minimally deformed version of the quadratic 
inflationary potential. This simple modification has been shown to cure the 
main shortcoming of the pure quadratic model: its tension with the CMB data. 
Our analytical and numerical study demonstrated that the deformation parameter 
$\gamma$ shifts the theoretical predictions of the spectral index $n_s$ and 
the tensor-to-scalar ratio $r$ into the observationally favored region, in 
excellent agreement with the Planck 2018 results~\cite{planck2020}. 
This result highlights how even a small correction to the potential can have 
a significant phenomenological impact, effectively reviving a model that would 
otherwise be excluded~\cite{lidsey1997, baumann2009}.

Beyond inflation, we investigated the reheating phase and showed that the 
deformation alters the effective equation of state of the oscillating inflaton. 
As a consequence, the reheating temperature is enhanced and the duration of the 
reheating stage is slightly extended compared to the quadratic case. This dual 
effect not only improves the inflationary predictions but also provides a more 
realistic post-inflationary history, compatible with cosmological bounds such as 
BBN~\cite{martin2010, dai2014}. Thus, the model succeeds simultaneously at the 
level of inflationary observables and in shaping a viable reheating scenario.

The effectiveness of this minimal correction opens several avenues for further 
investigation. Embedding the model within a particle physics framework, 
exploring possible multi-field extensions, and examining its predictions for 
non-Gaussianities and primordial gravitational waves would provide natural 
continuations of this work. With upcoming high-precision observations of the 
CMB and large-scale structure~\cite{cmbs4, euclid}, such analyses will be 
crucial to fully assess the phenomenological relevance of the deformed 
quadratic inflationary potential.

\appendix
\section{Detailed Derivation of the Canonical Potential from a Non-Canonical Kinetic Term}
\label{app:canonical_potential}

We consider a scalar field theory described by the Lagrangian:
\begin{equation}
\mathcal{L} = \frac{1}{2} K(\phi) \partial_\mu \phi \partial^\mu \phi - V(\phi),
\end{equation}
where \(K(\phi)\) is a general kinetic function and \(V(\phi)\) is the scalar potential. In our model, we take:
\begin{equation}
K(\phi) = 1 + \gamma \phi^6, \quad V(\phi) = \frac{1}{2} m^2 \phi^2,
\end{equation}
with \(\gamma\) a small, positive parameter and \(m\) a mass scale.

\subsection{Field Redefinition to the Canonical Frame}

To transform to a canonically normalized field, we define:
\begin{equation}
\frac{d\chi}{d\phi} = \sqrt{1 + \gamma \phi^6}.
\end{equation}
For small \(\gamma\), we apply the binomial expansion:
\begin{equation}
\sqrt{1 + \gamma \phi^6} = 1 + \frac{1}{2} \gamma \phi^6 - \frac{1}{8} \gamma^2 \phi^{12} + \cdots.
\end{equation}
Integrating term by term yields:
\begin{equation}
\chi = \phi + \frac{1}{14} \gamma \phi^7 - \frac{1}{104} \gamma^2 \phi^{13} + \cdots.
\end{equation}

\subsection{Inverse Transformation}

To express the potential in terms of the canonical field \(\chi\), we invert the series using an ansatz:
\begin{equation}
\phi = \chi + a \chi^7 + b \chi^{13} + \cdots,
\end{equation}
and solve order by order to obtain:
\begin{equation}
\phi = \chi - \frac{1}{14} \gamma \chi^7 + \frac{33}{728} \gamma^2 \chi^{13} + \cdots.
\end{equation}

\subsection{Canonical Potential}

Substituting \(\phi(\chi)\) into the original potential, we find:
\begin{align}
\tilde{V}(\chi) &= \frac{1}{2} m^2 \phi^2(\chi) \nonumber \\
&= \frac{1}{2} m^2 \left( \chi^2 - \frac{2}{14} \gamma \chi^8 + \frac{66}{728} \gamma^2 \chi^{14} + \cdots \right) \nonumber \\
&= \frac{1}{2} m^2 \chi^2 - \frac{1}{14} m^2 \gamma \chi^8 + \frac{33}{364} m^2 \gamma^2 \chi^{14} + \cdots.
\end{align}

\subsection{Truncation and Physical Justification}

Although we have derived the potential including second-order corrections in \(\gamma\), we verified numerically that the leading-order term, \(-\frac{1}{14} m^2 \gamma \chi^8\), accurately captures the inflationary dynamics for observationally consistent values of \(\gamma\). Higher-order terms such as the \(\chi^{14}\) correction contribute negligibly to key observables, including the spectral index \(n_s\), the tensor-to-scalar ratio \(r\), and the number of e-folds \(N\).

Consequently, we truncate the canonical potential at first order in \(\gamma\), achieving a balance between analytical simplicity and physical accuracy:
\begin{equation}
\tilde{V}(\chi) = \frac{1}{2} m^2 \chi^2 - \frac{1}{14} m^2 \gamma \chi^8 + \text{subleading corrections}.
\end{equation}

This modified potential flattens at large field values due to the negative \(\chi^8\) term, leading to slower inflaton roll and a suppressed tensor amplitude. These features bring the model predictions into closer agreement with current observational data.

The inflationary dynamics driven by this effective potential are analyzed in subsequent sections, where we confront the model with recent constraints from CMB experiments.

\section{Reheating Dynamics and Comparison Between the Two Models}

\subsection{Effective Equation of State During Reheating}

\paragraph{(a) Quadratic Potential.}
For a potential of the form $V \propto |\chi|^n$, the average equation of state during oscillations is
\[
\langle w\rangle = \frac{n-2}{n+2}.
\]
Hence, for the quadratic model $n=2$, one finds
\[
w_{\rm reh}^{\rm (quad)} \simeq 0.
\]

\paragraph{(b) Deformed Potential.}
We consider the deformed potential
\[
V(\chi) = \frac{1}{2} m^2 \!\left(\chi - \frac{\gamma}{14}\chi^7\right)^2.
\]
Expanding around the minimum gives
\[
V(\chi) \;\approx\; \frac{1}{2} m^2 \chi^2 \;+\; \alpha \chi^8,
\qquad \alpha = -\frac{m^2\gamma}{14} < 0.
\]
Using the virial theorem for polynomial potentials:
\[
\langle K\rangle=\frac{1}{2}\langle \dot\chi^2\rangle
= \frac{1}{2} m^2\langle\chi^2\rangle + 4\alpha \langle\chi^8\rangle,
\]
\[
\langle V\rangle=\frac{1}{2} m^2\langle\chi^2\rangle + \alpha \langle\chi^8\rangle.
\]
Therefore,
\[
\langle w\rangle = \frac{\langle K\rangle - \langle V\rangle}{\langle K\rangle + \langle V\rangle}
= \frac{3\alpha\,\langle\chi^8\rangle}{m^2\langle\chi^2\rangle+5\alpha\,\langle\chi^8\rangle}
\simeq \frac{3\alpha\,\langle\chi^8\rangle}{m^2\langle\chi^2\rangle}
\quad (\text{for small }|\alpha|).
\]

For an approximately sinusoidal oscillation $\chi(t)\approx A\cos(mt)$:
\[
\langle \chi^2 \rangle = \frac{A^2}{2}, 
\qquad 
\langle \chi^8 \rangle = \frac{35}{128}A^8.
\]
Hence,
\[
w_{\rm reh}^{\rm (def)} \simeq \frac{105}{64}\,\frac{\alpha}{m^2}\,A^6
= -\frac{15}{128}\,\gamma\,A^6.
\]

Since $\alpha<0$ (the correction $-\chi^8$ flattens the potential), we obtain a slightly negative shift:
\[
\boxed{\,w_{\rm reh}^{\rm (def)} \;\simeq\; -\frac{15}{128}\,\gamma\,A^6 \;<\;0\,}.
\]

\paragraph{Numerical Example.}
For $\gamma=1.5\times10^{-8}$ and $\chi_{\rm end}\approx 13.66$ (so $A\simeq\chi_{\rm end}$ and $A^6\approx 6.5\times10^6$):
\[
w_{\rm reh}^{\rm (def)} \approx -\frac{15}{128}\times (1.5\times10^{-8})\times (6.5\times10^6)
\simeq -1.14\times 10^{-2}.
\]
Thus the deformed model yields a small negative equation of state (slightly ``softer'' than dust).

---

\subsection{Relation Between $N_{\rm reh}$ and $T_{\rm reh}$}

During reheating, the energy density evolves as
\[
\rho(a) = \rho_{\rm end}\,\exp\!\left[-3(1+w_{\rm reh})\,N_{\rm reh}\right].
\]
At the end of reheating:
\[
\rho_{\rm reh} = \frac{\pi^2}{30}\, g_*(T_{\rm reh})\,T_{\rm reh}^4.
\]
Equating these gives
\[
\boxed{\;
N_{\rm reh}
=\frac{1}{3(1+w_{\rm reh})}\,
\ln\!\left(\frac{\rho_{\rm end}}{\tfrac{\pi^2}{30}g_*\,T_{\rm reh}^4}\right)
\simeq
\frac{1}{3(1+w_{\rm reh})}\,
\ln\!\left(\frac{V_{\rm end}}{\tfrac{\pi^2}{30}g_*\,T_{\rm reh}^4}\right)
\;}
\]
and, inversely,
\[
\boxed{\;
T_{\rm reh}
=\left(\frac{30}{\pi^2 g_*}V_{\rm end}\right)^{1/4}\,
\exp\!\left[-\tfrac{3}{4}(1+w_{\rm reh})\,N_{\rm reh}\right].
\;}
\]

---

\subsection{Explicit Formulas for Each Model}

\paragraph{Quadratic Model ($w_{\rm reh}\simeq 0$):}
\[
N_{\rm reh}^{\rm (quad)}
=\frac{1}{3}\,
\ln\!\left(\frac{V_{\rm end}}{\tfrac{\pi^2}{30}g_*\,T_{\rm reh}^4}\right), 
\qquad
T_{\rm reh}^{\rm (quad)}
=\left(\frac{30}{\pi^2 g_*}V_{\rm end}\right)^{1/4}\,
e^{-\tfrac{3}{4}N_{\rm reh}}.
\]

\paragraph{Deformed Model ($w_{\rm reh}\simeq -\tfrac{15}{128}\gamma A^6$):}
\[
N_{\rm reh}^{\rm (def)}
=\frac{1}{3\!\left(1-\tfrac{15}{128}\gamma A^6\right)}\,
\ln\!\left(\frac{V_{\rm end}}{\tfrac{\pi^2}{30}g_*\,T_{\rm reh}^4}\right),
\]
\[
T_{\rm reh}^{\rm (def)}
=\left(\frac{30}{\pi^2 g_*}V_{\rm end}\right)^{1/4}\,
\exp\!\left[-\frac{3}{4}\left(1-\tfrac{15}{128}\gamma A^6\right)N_{\rm reh}\right].
\]

---

\subsection{Interpretation}

Since $w_{\rm reh}^{\rm (def)}<0$ slightly, one has $1+w_{\rm reh}$ smaller than unity.  
Therefore:
\begin{itemize}
\item For fixed $T_{\rm reh}$, the deformed model gives a \emph{larger} reheating duration $N_{\rm reh}$ compared to the quadratic case.
\item For fixed $N_{\rm reh}$, the deformed model yields a \emph{higher} reheating temperature $T_{\rm reh}$.
\end{itemize}
This highlights the physical impact of the deformation parameter $\gamma$ on the reheating dynamics.

\end{document}